\documentclass[journal]{IEEEtran}
\usepackage{changes}
\usepackage{filecontents} 
\usepackage{cite}
\usepackage{graphicx}
\usepackage{xcolor}

\ifCLASSINFOpdf
\else

\fi

\usepackage{array}

\usepackage{graphicx}
\usepackage{epsf}
\usepackage{float}
\usepackage{stfloats}
\usepackage{amssymb}
\begin{document}
\bstctlcite{IEEEexample:BSTcontrol}

\title{Target Flow-Pressure Operating Range for Designing a Failing Fontan Cavopulmonary Support Device}

\author{Masoud~Farahmand,~\IEEEmembership{}
        Minoo N.~Kavarana,~\IEEEmembership{}
        Phillip M.~Trusty,~\IEEEmembership{}
        and~Ethan O.~Kung~\IEEEmembership{}
\thanks{This work was supported by Clemson University, an award from the American Heart association and The Children’s Heart Foundation (16SDG29850012) and an award from the National Science Foundation (1749017). M. Farahmand (mfarahm@g.clemson.edu) is with Clemson University, Clemson, SC, USA; M. N. Kavarana (kavarana@musc.edu) is with the Medical University of South Carolina, Charleston, SC, USA; P. M. Trusty (phillip.trusty@gmail.com ) is with Georgia Institute of Technology; E. O. Kung (correspondence email: ekung@clemson.edu) is with Clemson University, Clemson, SC, USA.}
\thanks{Manuscript received July 01, 2019; revised December 16, 2019}
\thanks{ }
\thanks{Copyright (c) 2020 IEEE. Personal use of this material is permitted.
However, permission to use this material for any other purposes must
be obtained from the IEEE by sending an email to
pubs-permissions@ieee.org.}}

\maketitle

\begin{abstract}
Fontan operation as the current standard of care for the palliation of single ventricle defects results in significant late complications.
Using a mechanical circulatory device for the right circulation to serve the function of the missing subpulmonary ventricle could potentially stabilize the failing Fontan circulation.
This study aims to elucidate the hydraulic operating regions that should be targeted for designing cavopulmonary blood pumps.
By integrating numerical analysis and available clinical information, the interaction of the cavopulmonary support via the IVC and full assist configurations with a wide range of simulated adult failing scenarios was investigated; with IVC and full assist corresponding to the inferior venous return or the entire venous return, respectively, being routed through the device. 
We identified the desired hydraulic operating regions for a cavopulmonary assist device by clustering all head pressures and corresponding pump flows that result in hemodynamic improvement for each simulated failing Fontan physiology. 
Results show that IVC support can produce beneficial hemodynamics in only a small fraction of failing Fontan scenarios. Cavopulmonary assist device could increase cardiac index by
35\% and decrease the inferior vena cava pressure by 45\% depending
on the patient’s pre-support hemodynamic state and surgical
configuration of the cavopulmonary assist device (IVC or full
support). The desired flow-pressure operating regions we identified can serve as the performance criteria for designing cavopulmonary assist devices as well as evaluating off-label use of commercially available left-side blood pumps for failing Fontan cavopulmonary support.
\end{abstract}

\IEEEpeerreviewmaketitle
\section{Introduction}
\IEEEPARstart{S}{ingle} ventricle palliation involving three stages of surgeries has become an accepted common practice over the past 50 years~\cite{davey2019surveillance}. The Fontan operation involves diverting the entire systemic venous return directly to the pulmonary arterial tree most commonly via the formation of the total cavopulmonary connection (TCPC), where most surgeons use an extracardiac conduit to bypass the atrium completely~\cite{van2018state}.

While the Fontan procedure has resulted in survival of single ventricle patients into adulthood and reduced early morbidity and mortality, the lack of the subpulmonary ventricle to drive the flow forward into the lungs and the pressure loss across the TCPC (usually 2-5mmHg~\cite{rodefeld2011cavopulmonary}) have led to a myriad of serious late complications for these patients. For instance, excessive hepatic venous pressure in these patients leads to complications such as hepatic cirrhosis, hepatic fibrosis, and eventual liver failure~\cite{kuwabara2018liver}.

A significant number of patients following the Fontan operation (post-operation$>$15 years; age: 23$\pm$7 years~\cite{ohuchi2017hemodynamic}) ultimately develop circulatory failure. The failing stage in these patients is traditionally characterized by significant hemodynamic anomalies such as high caval pressure, low cardiac output, and low arterial oxygen saturation~\cite{GEWILLIG2014105,ohuchih2014comparisonof}. Recently, in addition to the traditional class of failing Fontan patients, a new failing Fontan phenotype class with high caval pressure and normal cardiac output has been recognized~\cite{book2016clinical,ohuchi2017optimal}. Despite normal cardiac output, this failing Fontan class is sicker compared to the traditional group, possibly due to more severe liver-related complications~\cite{miranda2019haemodynamic}. While some researchers have found no association between Fontan related liver complications and cardiac output~\cite{goldberg2017hepatic}, others have suggested liver fibrosis and hepatic arterialization as a cause for the higher cardiac output in this group of failing Fontan patients~\cite{trusty2018impact}. 

Even though cardiac transplantation is considered throughout the patient$'$s lifetime, many of Fontan patients are not able to receive transplantation as a result of several previous surgeries and multiple organ dysfunctions as well as unavailability of matching organ donors. The growing population of Fontan patients suffering from life-threatening late complications signifies the importance of undertaking engineering and medical efforts to address the unmet therapeutic need and ameliorate limitations of the Fontan circulation.

In 1998, de Leval~\cite{de1998fontan} proposed the use of a mechanical cavopulmonary assist device to target the cavopulmonary segment of the circulation. This proposed device is expected to substitute the function of the absent subpulmonary ventricle as a solution for reducing the caval pressure and augmenting the ventricular filling in a failing Fontan circulation. Such a device should not be obstructive in the event of operational failure. However, there is currently no pump commercially available for the cavopulmonary application and right side support is not used in the clinical management of Fontan patients. 

Cavopulmonary support for failing Fontan patients has been unsuccessful as a result of different hemodynamic conditions between these patients~\cite{Trusty2019}. A suitably designed cavopulmonary assist device should have performance profiles versatile enough to serve a wide range of failing Fontan patients with various levels of hemodynamic instability. The objective of this study is to identify the desired operating regions for such a device in the setting of inferior vena cava (IVC) and full assist surgical configurations.

In this study, we classified hemodynamic failure in Fontan patients into two groups: 1) traditional failure with low cardiac output concurrent with chronic high caval pressure and 2) failure with chronic high caval pressure concurrent with normal cardiac output. Each group encompasses physiologic scenarios with various hemodynamic states. We developed numerical models to simulate a full range of failing stage Fontan physiologies in each phenotype class reflecting pathological conditions with various severity levels of systolic dysfunction, diastolic dysfunction, and abnormal systemic and pulmonary vascular resistances. The Fontan physiologic scenarios with different hemodynamic conditions were modeled to represent approximately 95\% of patients in both failing phenotype groups. Next, a cavopulmonary assist device generating a range of head pressure rises (from 0-20mmHg) during full and IVC support was modeled for each physiologic scenario. In each simulated scenario, we identified the pump head pressures and the resulting pump flows (P-Q data points) that produce hemodynamic improvement, signified by reduced IVC pressure (most important) and other desirable hemodynamic parameters. Lastly, we identified the desired operating regions for designing a cavopulmonary assist device suitable for the full or IVC support by clustering all of the P-Q data points that resulted in hemodynamic improvement in all simulated failing Fontan physiology.

\section{Methods}
\subsection{Physiology model of the functional Fontan circulation (Baseline)}
\begin{figure}[tb!]
    \centerline{\includegraphics[width=88mm]{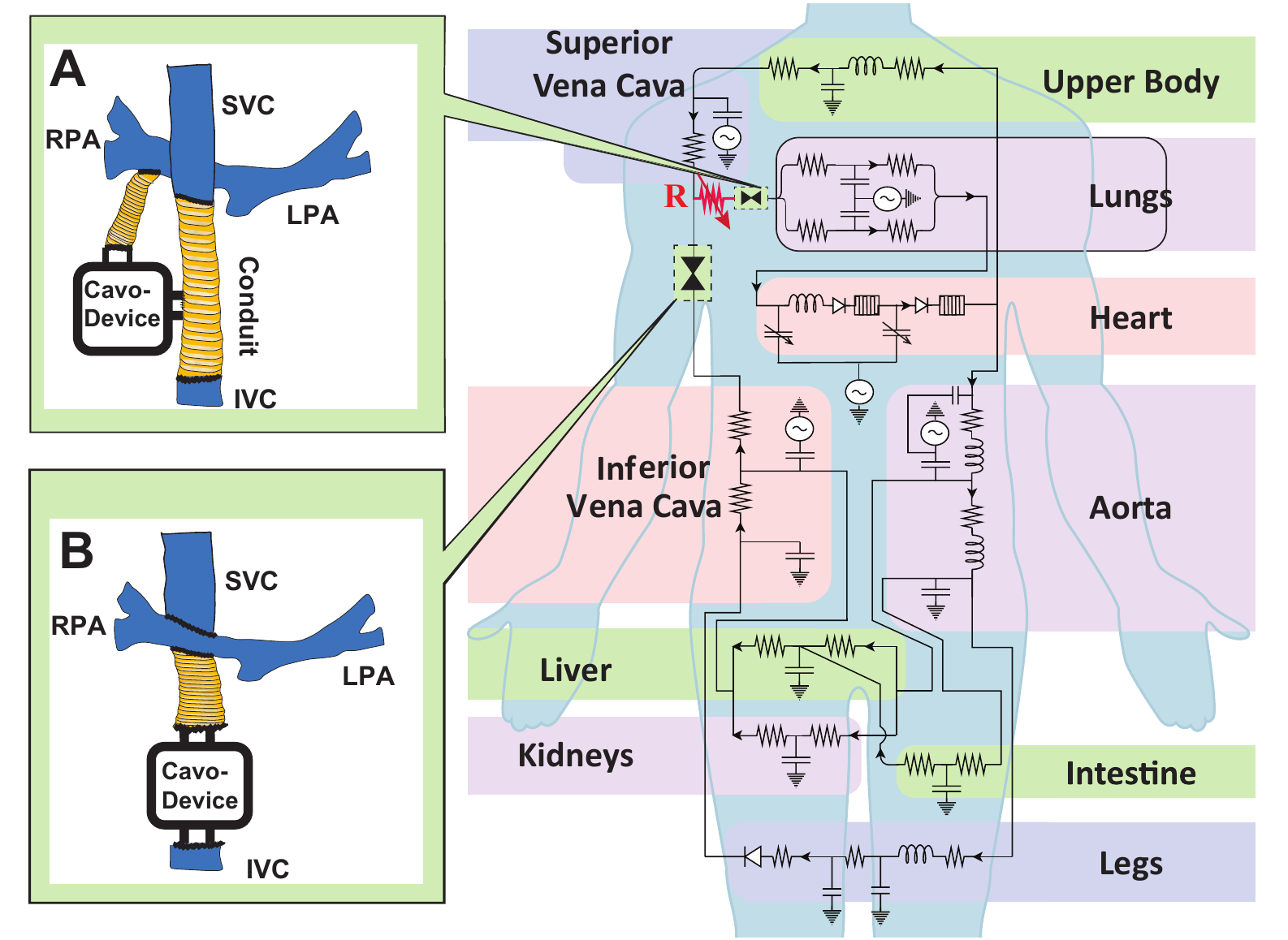}}
    \caption{\textbf{Closed-loop lumped parameter model of the Fontan circulation}. Conceptual representation of the (\textbf{A}) Full support and (\textbf{B}) IVC support assist device coupled to the LPN model of the Fontan circulation. \textbf{\textcolor{red}{R}} is the cavopulmonary connection dynamic resistance, \textbf{Cavo-Device} is cavopulmonary assist device, \textbf{LPA} is left pulmonary artery, \textbf{RPA} is right pulmonary artery \textbf{SVC} is superior vena cava, \textbf{IVC} is inferior vena cava. Pressure points ($P_{sub}$) are labeled on the diagram. $L_{sub}$ and $R_{sub}$ are inductor and resistor components, respectively~\cite{kung2014simulation}.}
     \label{fig1}
\end{figure}

In an earlier study, we used clinical and physiological data from Fontan patients to develop and validate a closed-loop lumped parameter network (LPN) model for describing the functional Fontan circulation (Fig.~\ref{fig1}). The computational model is tuned based on patient weight and height and prescribes the cardiac function and pulmonary and systemic vascular impedances according to the patient’s body size. It was shown that the physiology model produces accurate trends in physiological parameters~\cite{kung2014simulation}. The heart block produces the pressure needed for driving the flow through the circuit. The cardiac function for generating the transmyocardial pressure is described using equations (1) and (2). E(t) is the ventricular time-dependent elastance; En(t) is the normalized elastance function; $E_{max}$, and $E_{offset}$ are constants related to contractility (as well as ventricular ejection fraction) and ventricular filling (and ventricular stiffness), respectively. t describes the time point in the cardiac cycle period and $t_{svs}$ is the cardiac systolic period. Ventricular volume and reference volume are $V_{sv}$ and $V_{sv0}$, respectively. The left and right pulmonary artery flow split is assumed to be 50/50. 

\begin{equation} \label{eq1}
P_{sv}(V_{sv},t)= E(t)(V_{sv}-V_{sv0})
\end{equation}
\begin{equation} \label{eq2}
E(t)= E_{max}En(\frac{0.3}{t_{svs}}t)+E_{offset}
\end{equation}
Table~\ref{tab1} details the range of time-averaged values of selected physiologic parameters of six Fontan patients (baseline) with various body sizes. The simulation are performed based on the assumption that these six body sizes cover a broad range of Fontan patients. Parameters used for setting the LPN component values for each patient at the baseline are derived according to our previous study~\cite{kung2014simulation}.

\begin{table}[tb!]
\centering
\caption{Mean values of the hemodynamic parameters from simulations of the 6 example Fontan patients at baseline.}
\begin{tabular}{ll}
\hline
Parameters                & Range             \\ \hline
BSA (m$^2$)                  & {[}1.27-2.3{]}    \\
Cardiac output (L.min$^{-1}$)  & {[}3.37-5.79{]}   \\
Arterial pressure(mmHg)   & {[}91.98-97.10{]} \\
Atrial pressure(mmHg)     & {[}8.90-9.92{]}   \\
Pulmonary pressure (mmHg) & {[}13.37-14.50{]} \\
IVC pressure (mmHg)       & {[}13.37-14.50{]} \\ 
\hline
\end{tabular}
\label{tab1}
\end{table}
\subsection{Physiology model of patients with hemodynamic failure}
 In this study, we adjusted parameter values in the baseline LPN model to simulate a wide range of Fontan failure. There are several determinants which influence hemodynamic parameters such as patient's body size, cardiac function and the aberration level of systemic and pulmonary vascular resistances.  The baseline model (described in previous section) is used as a starting point for building physiology models of patients in the failing stage. This was achieved by adjusting the component values corresponding to the pulmonary and systemic vascular resistances as well as the systolic and diastolic ventricular functions. To include a wide range of patient scenarios for each phenotype group, for each of the six patient body size baseline models, we simulated different physiologic scenarios with different level of pulmonary and systemic vascular resistances and severity of systolic and diastolic dysfunctions. The structure of the LPN circuit is not affected by these changes, only the component values are. 
Ohuchi et al. ~\cite{ohuchi2017hemodynamic} and Cavalcanti et al.~\cite{cavalcanti2001analysis} have evaluated the hemodynamics in adult failing Fontan patients and presented a broad range of clinically measured physiologic values of cardiac index and arterial and IVC pressures for failing patients as mean value with 1 standard deviation (SD). We modeled physiologic scenarios in each phenotype class such that the range of simulated scenarios (e.g., cardiac index, arterial pressure, IVC pressure) cover approximately 2SD of those clinically measured parameter values; based on the three-sigma rule, 2SD from the mean value encompasses $\approx$95\% of the population values~\cite{bryc2012normal}.\\

\subsubsection{LPN Model of Group I phenotype (Failing patients with high caval pressure and low cardiac output)}
This phenotype is well recognized. The circulation in this class of patients generally deteriorates as a result of systolic dysfunction, diastolic dysfunction, and high pulmonary vascular resistance~\cite{ohuchi2017hemodynamic,ohuchi2017optimal}. Therefore, we simulated this phenotype by reducing ventricular contractility ($E_{max}$, representing ventricular systolic function), increasing ventricular stiffness ($E_{offset}$, representing ventricular diastolic function) and increasing pulmonary vascular resistance (PVR). These values were incrementally changed to model various severity levels of hemodynamic aberration according to the following protocol:\\ \\
\textbf{1)}	The range of $E_{max}$ was defined to start at the baseline value and end at the value that results in an ejection fraction of 30\%~\cite{di2015simulation} (this corresponds to a specific cardiac output).\\
\textbf{2)}	While $E_{max}$ was held constant at its baseline value, the range of $E_{offset}$ was defined to start at the baseline value and end at the value that results in the cardiac output obtained at the end of step (1).\\ 
\textbf{3)}	Similarly, while $E_{max}$ and $E_{offset}$ were maintained at their baseline values, PVR\'{s} range was defined to start at its baseline value and end at the value that results in the cardiac output obtained at the end of step (1).\\
\textbf{4)}	 A number of equally spaced levels for each input variable (PVR, $E_{max}$, $E_{offset}$) over their ranges, as determined in steps (1-3), were used in all possible permutations to simulate a wide range of possible failing physiologic scenarios. \\
\textbf{5)} Next, selected physiologic parameters (IVC pressure, cardiac index, and arterial pressure) from each simulated scenario were evaluated against clinical catheterization data from failing Fontan patients reported by Cavalcanti et al.~\cite{cavalcanti2001analysis}.\\The scenarios where the selected physiologic parameters fall outside of approximately 2SD of the mean value of the clinical catheterization data were discarded. This resulted in a final total of 630 simulated patient scenarios for group I. \\
\subsubsection{LPN Model of Group II phenotype (Failing patients with high caval pressure and normal cardiac output)}
To our knowledge, there exists currently no mathematical circulation model available for this group. The pathophysiology related to this class of patients is not well understood. While the cardiac function in this group of patients is normal, the systemic vascular resistance (SVR) is abnormally low. Moreover, contrary to the traditional group of failing Fontan patients, the pulmonary vascular resistance in these patients is often low~\cite{ohuchi2017optimal}. To simulate this class of Fontan failures, we decreased the systemic and pulmonary vascular resistances over a wide region from their baseline values while maintaining normal ventricular systolic and diastolic function. There is limited information on the level of systemic and pulmonary vascular resistance aberration in this recently recognized group of failing patients. Therefore, we investigated the SVR and PVR values in our models such that the selected physiologic parameters from simulations cover 2SD of the mean value of the clinical data through the following steps:\\ \\
\textbf{1)}	The region of interest of SVR was defined to start from its baseline value (SVR$_0$) and end at 10\% of SVR$_0$. 10 equally spaced levels for SVR over its region of interest were taken.\\ 
\textbf{2)}	The region of interest of PVR was defined to start from its baseline value (PVR$_0$) and end at 20\% of PVR$_0$ . 10 equally spaced levels for PVR over its region of interest were taken.\\
\textbf{3)}	Similar to the procedure used for group I, we compared the simulated physiologic scenarios in group II with clinical data from group II failing patients reported by Ohuchi et al.~\cite{ohuchi2017hemodynamic} and removed physiologic scenarios that were outside of 2SD of the mean value of the clinical data. These steps resulted in a final total of 334 simulated patient scenarios for group II.
\subsection{Cavopulmonary assist device implementation (Fig.~\ref{fig1})} 
We modeled the implementation of the cavopulmonary support via IVC and full assist surgical configurations. The graphic in Fig.~\ref{fig1} is for the purpose of illustrating the blood routing concept of the IVC/full support configurations, not as a geometric surgical guide for device placement. In a realistic clinical scenario, the implant will most likely be extracorporeal.

The dynamic flow resistance of the surgical junction geometry is flow-related; we therefore included a variable resistance in the LPN model as shown in Fig.~\ref{fig1} to estimate the dynamic changes in the surgical junction resistance as a result of pump-induced flow augmentation. The details of the experimental setup for characterizing the value of this dynamic resistance are included in the Appendix A.
\subsubsection{Model of full assist configuration (Fig.~\ref{fig1}A)}
In this configuration, the pump is implemented such that it drives the entire venous return to the lungs. Implanting the pump would require performing a Fontan ”takedown” and additional surgeries for anastomosis of the superior vena cava (SVC) and IVC conduit; this resulted in both vena cavae being at the pump’s inlet and, having the same blood pressure. For the purpose of illustration, Fig.~\ref{fig1}A shows a potential configuration where an extra conduit connects the cavopulmonary device between the vena cava and the pulmonary artery.
\subsubsection{Model of IVC assist configuration (Fig.~\ref{fig1}B)}
In this configuration only the IVC flow passes through the pump. The IVC pump is implemented in series with the Fontan baffle and the pump's inlet and outlet are connected to the IVC and the right/left pulmonary arteries confluence in the LPN circuit, respectively. The zero-dimensional LPN model does not contain spatial information, therefore, the TCPC junction being at the same node as the SVC shares an equal blood pressure; and the IVC shares the same pressure as the pump’s inlet.

\subsection{Desired operating regions for cavopulmonary assist device}
Overall, 964 (630 for group I + 334 for group II) different patient scenarios were simulated. For each patient scenario
and with each the full and IVC assist surgical configurations, we simulated device head pressure rises from 0-20mmHg in 0.5mmHg increments. Overall, approximately 77,000 (964 patients scenarios$\times$2 configurations$\times$40 head pressure rise values) simulations were carried out to obtain the desired operating regions. The algebraic and ordinary differential equations obtained from the LPN models were solved using Runge Kutta method programmed in Fortran. Simulations were carried out for 30 seconds to ensure that all flows and pressures in the LPN circuit stabilize. All results we report in this paper reflect the stabilized state.

High caval pressure is the main determinant of Fontan related diseases. Thus, for each simulated scenario, the head pressures and the resultant corresponding pump flows that led to a decrease in the caval pressure while maintaining other physiologic pressures in safe ranges were identified as favorable. As such, hemodynamic improvement was defined as at least 3mmHg decrease in the IVC pressure while maintaining a pulmonary pressure $<$25mmHg~\cite{montani2013pulmonary} (to avoid pulmonary arterial hypertension and lung perfusion damage), an atrial pressure of $<$16mmHg~\cite{larue2017clinical}, SVC pressure $<$20mmHg (to avoid precipitous increase in the SVC pressure and potential cerebral damage), and mean IVC pressure $>$4mmHg (to avoid potential IVC collapse). 
Finally, the data points corresponding to the head pressures and pump flows that resulted in hemodynamic improvement in each physiologic scenario were clustered to form the desired operating regions. The flow diagram (Fig.~\ref{figA1}) in Appendix B summarizes the process for identifying the desired operating regions.

\section{Results}
\subsection{Patients with hemodynamic failure}
\begin{figure}
    \centerline{ \includegraphics[width=88mm]{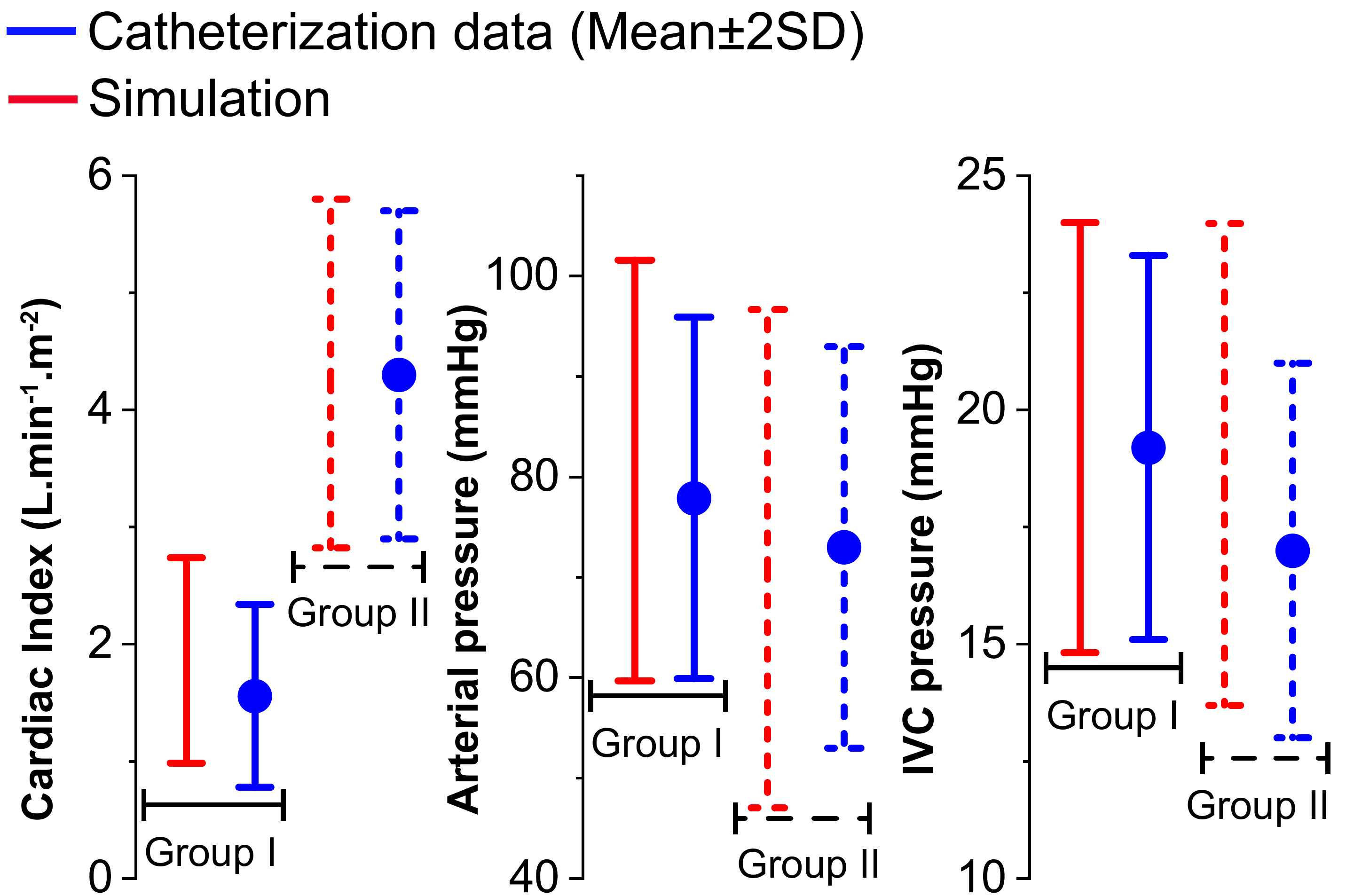}}
    \caption{\textbf{Hemodynamics of failing Fontan patients from simulations vs. clinical data}. Range of the selected hemodynamic parameters (pre-support) from simulations and reported catheterization data~\cite{cavalcanti2001analysis,ohuchi2017hemodynamic} of the group I and II failing Fontan physiologic scenarios. CI: cardiac index, IVC pressure: inferior vena caval pressure, SD: Standard deviation.}
         \label{fig2}
\end{figure}

Comparing the simulation results against the catheterization information reported in Cavalcanti et al.~\cite{cavalcanti2001analysis} for group I and Ohuchi et al.~\cite{ohuchi2017hemodynamic} for group II of Fontan failures, the range of selected physiologic parameters for the simulated scenarios encompasses approximately a 2SD range of the values of those parameters measured clinically in Fontan failures in group I and II (Fig.~\ref{fig2}).

\subsection{Desired operating regions for cavopulmonary assist device}
\begin{figure}[tb!]
    \centerline{ \includegraphics[width=88mm]{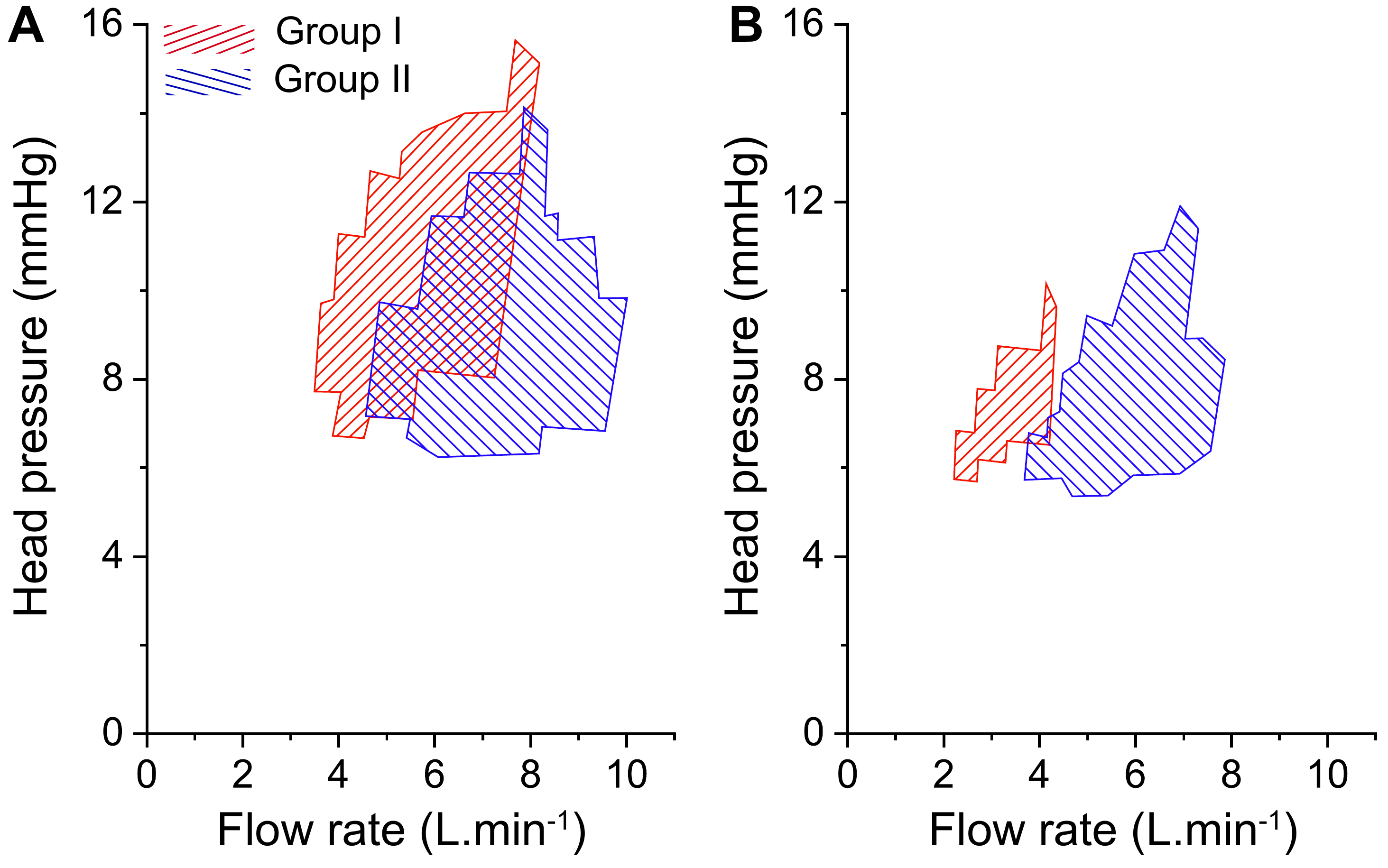}}
    \caption{\textbf{Target flow-pressure regions for designing cavopulmonary supports.} Desired operating regions of (\textbf{A}) full and (\textbf{B}) IVC assist cavopulmonary devices suitable for serving a wide range of failing Fontan patients in group I and II phenotype classes.}
         \label{fig3}
\end{figure}

We obtained the desired head pressure–flow operating regions of the IVC and full cavopulmonary assist supports that can serve a wide range of failing Fontan patients in each phenotype class (Fig.~\ref{fig3}). Depending on the patient’s class (group I or group II) and surgical configuration for installation of the cavopulmonary assist, a device should be designed such that its characteristic curves at different pump speeds can encompass the entire desired operating region.

\begin{figure*}[tb!]
    \centerline{ \includegraphics[width=130mm]{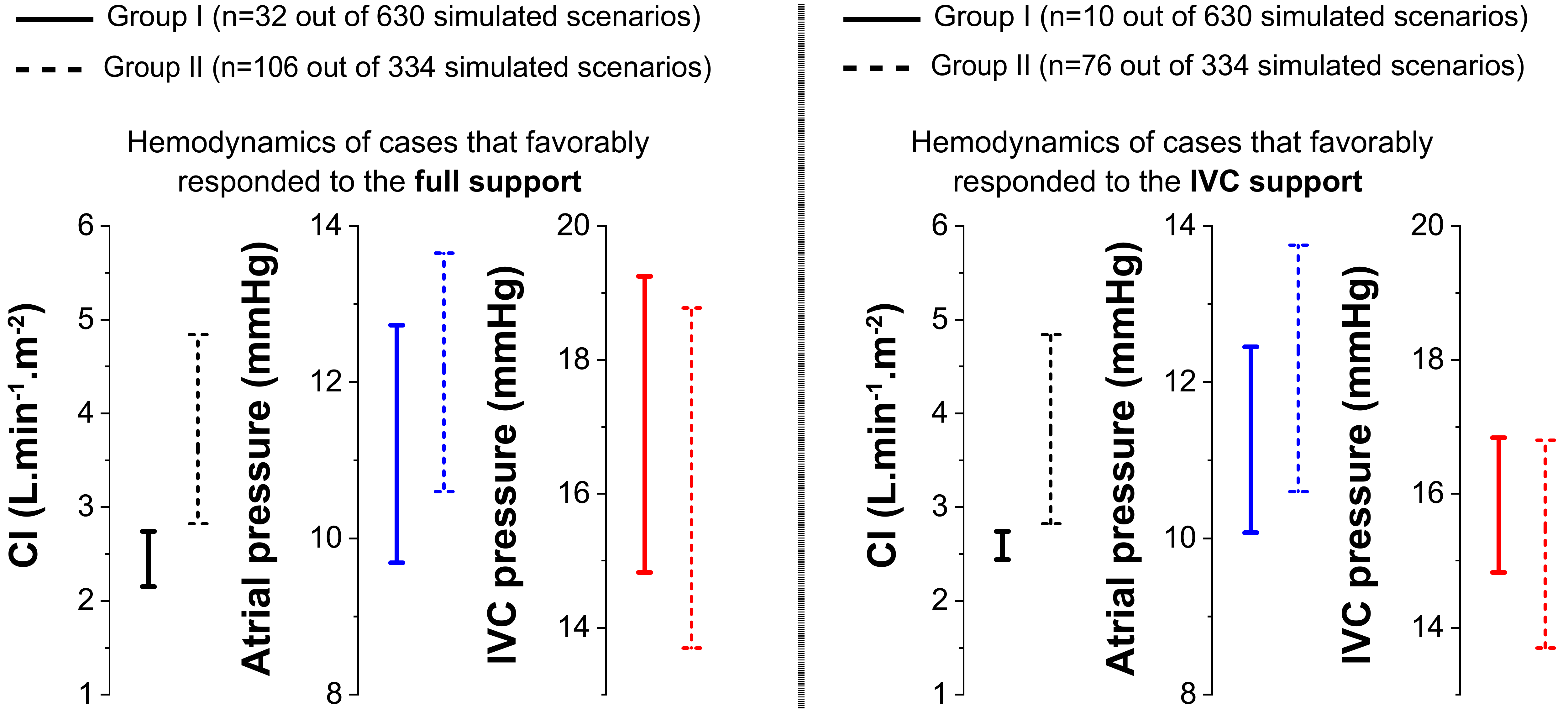}}
    \caption{\textbf{Pre-support hemodynamics of simulated patient scenarios that favorably
 responded to cavopulmonary supports.} Range of selected hemodynamic parameters (pre-support) of physiologic scenarios characterized as group I (Solid lines) and group II (Dashed lines) that favorably responded to the IVC and full cavopulmonary support.}
         \label{fig4}
\end{figure*}

We identified the simulated physiologic scenarios in each phenotype group that the presence of the cavopulmonary pump can lead to hemodynamic improvement (Fig.~\ref{fig4}). Full support did not benefit group I and group II if the patient’s pre-support IVC pressure $\gtrapprox$19.5mmHg and $\gtrapprox$18.5mmHg, respectively. Also, IVC support did not benefit group I and group II patients if the pre-support IVC pressure was $\gtrapprox$17mmHg. 
IVC support benefited significantly fewer physiologic scenarios comparing to the full support. For example, IVC support had beneficial hemodynamic outcome in only 10 simulated group I physiologic scenarios.

\subsection{Hemodynamic response to cavopulmonary support}
\begin{figure}[tb!]
    \centerline{ \includegraphics[width=88mm]{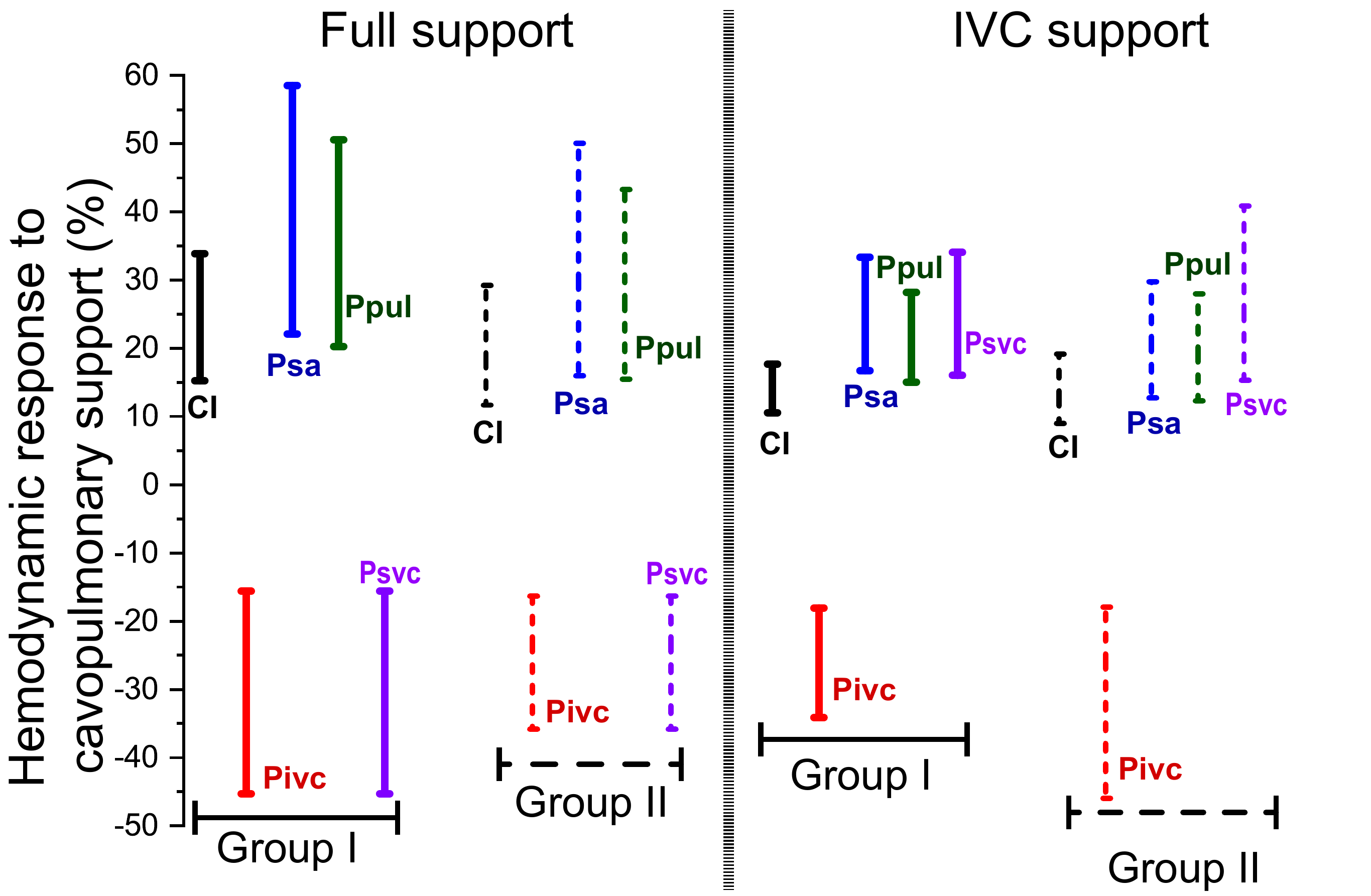}}
    \caption{\textbf{Cavopulmonary support hemodynamic impact}. Response range (\%) of the selected hemodynamic parameters to IVC and full cavopulmonary support. \textbf{CI}: Cardiac index; \textbf{P$_{ivc}$}: IVC pressure; \textbf{P$_{pul}$}: Pulmonary pressure; \textbf{P$_{sa}$}: Atrial pressure; \textbf{P$_{svc}$}: Superior vena cava pressure .}
         \label{fig5}
\end{figure}

Based on simulation results for the physiologic scenarios that favorably responded to the cavopulmonary support, the presence of the cavopulmonary assist promoted the cardiac index, pulmonary pressure and decreased the IVC pressure (Fig.~\ref{fig5}).

The cardiac index increase was 15-35\% (9-17\%) for group I and 10-28\% (7-19\%) for group II via full support (IVC support). The IVC pressure as a crucial parameter, also favorably decreased across both groups by the presence of the cavopulmonary assist. Overall, the maximum favorable reduction of the IVC pressure was $\approx$45\% via full support or IVC support. The IVC pump increased the SVC pressure, whereas the SVC pressure decreased via full support. 
\section{Discussion}
The present study aimed to identify the desired operating regions for a cavopulmonary blood pump that can support the failing Fontan circulation. These results can assist manufacturers in designing a right side assist device suitable for IVC or full support surgical configurations and for different classes of Fontan failures. To obtain these operating regions, a numerical analysis was performed investigating the interaction of a cavopulmonary assist device with a broad range of the possible failing Fontan physiologic scenarios.

In this study, cavopulmonary support in two groups of failing Fontan patients was investigated. These patients’ hemodynamic conditions depend on several factors, including systemic and pulmonary vascular resistances, and the cardiac performance which vary widely between these patients. Based on the studies by Ohuchi et al.~\cite{ohuchi2017hemodynamic} and Miranda et al.~\cite{miranda2019haemodynamic}, these two are major phenotypes of failing Fontan circulation. To our knowledge, each category’s prevalence among the failing Fontan population is not known. Both groups of the patients have high caval pressure and previous studies have shown that many of the Fontan related complications are linked to chronically high caval pressure~\cite{gewillig2016fontan}. However, the patients with a combination of normal cardiac index and high caval pressure have worse long-term survival in comparison with the traditional group based on clinical experience~\cite{miranda2019haemodynamic}. Liver injury is significantly correlated with high caval pressure and the abnormally low systemic vascular resistance in these patients is hypothesized to be due to liver complications. The reduced afterload in this group of patients significantly contributes to the cardiac index level. Even though these patients have normal cardiac output and do not need cardiac output augmentation, the presence of a cavopulmonary device can reduce the caval pressure and may alleviate liver dysfunction~\cite{pretre2008right}.The hypothesis that reversing the liver damage can potentially stabilize the systemic vascular resistance and cardiac output requires further investigation. 

Several studies have shown that a suitable cavopulmonary assist device could benefit the hemodynamics by decreasing the IVC pressure and promoting the cardiac output~\cite{broda2018progress,zhou2019avalonelite,lin2019computational,pekkan2018vitro}. In another computational study, Shimizu et al.~\cite{shimizu2016partial} also confirmed a reduction of the IVC pressure and increase of cardiac output by the presence of a cavopulmonary assist device in the IVC. Molfetta et al.~\cite{di2015simulation} simulated the physiology of 10 failing Fontan scenarios (group I) and quantified an average of 34\% increase in the cardiac output with an IVC support device. Compared to previous studies~\cite{di2015simulation,shimizu2016partial}, our results show a narrower range of cardiac output increase in group I scenarios with a maximum increase of 17\%; this is likely due to the flow-related surgical junction dynamic resistance that was not accounted for in previous studies. In the IVC assist configuration, the dynamic resistance of the TCPC junction resulted in significant pressure drop across the cavopulmonary connection and increase of the SVC pressure which limited the beneficial effects of the pump in the IVC support configuration. 

Currently, there exists no information in the literature about the dynamic change in the resistance of the full support surgical junction geometry as a result of the pump-induced flow augmentation. In this study, we applied the same dynamic resistance for both the IVC and full support surgical junctions (Appendix A).

While the IVC support surgical junction geometry has little room for variations, full support can adapt many implementation approaches with drastically different resulting geometries. Therefore, the surgical junction resistance in an actual full support implementation will largely depend on the specific geometry resulting from the implementation details. To use the results reported in this paper for different full support implementations, the user should estimate the pressure losses at different flow rates of the intended full support surgical junction geometry and compare them to those calculated from the equation in the Appendix A. Next, the regions presented in Fig.~\ref{fig3}. should be offset vertically at each flow rate according to any pressure differences found in the comparison analysis. Considering that the surgical junction pressure loss variation can be directly compensated with increased or decreased pump pressure output, the variation would not affect other results reported in this paper other than those in Fig.~\ref{fig3}. 

Our results show that both failing phenotype groups better responded to full support~(Fig.~\ref{fig5}). A bigger fraction of the failing Fontan scenarios can be served with full support geometries ~(Fig.~\ref{fig4}). %
Previous pioneering studies on short term Fontan animal models have highlighted the importance of designing a proper mechanical assist device for right side support by recognizing the problems such as IVC collapse associated with off-label use of a left ventricular assist device (VAD) for cavopulmonary application~\cite{riemer2005mechanical,rodefeld2004cavopulmonary}. For example, Riemer et al.~\cite{riemer2005mechanical} investigated the impact of the right support using a left VAD (Thoratec HeartMate II, Thoratec Corporation, Calif) in 8 sheep and observed decrease of the IVC flow as a result of IVC collapse and low IVC pressure (-0.25$\pm$0.48mmHg) in all animals. 
In an earlier study~\cite{farahmand2019risks}, it was shown that there are risks associated with using a commercially available left VAD for cavopulmonary support in single ventricle patients. The pressure level in the cavopulmonary pathway is approximately 90\% lower compared to the left side arterial environment; a left VAD typically generates a head pressure of $\approx$37 to 80mmHg for flow rates ranging approximately from 0.5 to 6L.min$^{-1}$. The high head pressure produced by a left VAD implanted on the right circulation may result in pulmonary arterial tree perfusion damage, central venous collapse~\cite{rodefeld2011cavopulmonary}, cerebral edema or venous thrombosis. 

We studied the two most clinically envisioned surgical configurations for the installation of the cavopulmonary assist device (IVC and full support)~\cite{shimizu2016partial} and identified a range where hemodynamic improvement is possible for each patient scenario via full or IVC support. Theoretically, the optimal operating point for each patient lies somewhere in this range; therefore a full or IVC cavopulmonary device should be capable of covering the corresponding range to provide maximal adjustment flexibility to clinicians for obtaining the optimal operating point for any patient. Simulations show that a proper cavopulmonary device for the traditional group of failing Fontan patients (group I) only needs to cover a small operating region. However, for the newly recognized phenotype group, the desired operating region covers a larger area that may complicate the design of a cavopulmonary device.

There are other possible configurations for installation of the cavopulmonary assist. For instance, Jagani et al.~\cite{jagani2019dual} proposed a dual-headed intravascular cavopulmonary pump for installation in the SVC and IVC. In this case, the entire venous return is pumped to the lungs and it is conceptually identical to the extracorporeal full support case. Cavopulmonary assistance in the SVC is not a suitable option for two reasons: 1) most late Fontan pathophysiology emerges from high IVC pressure (liver diseases, protein losing enteropathy,etc)~\cite{rodefeld2011cavopulmonary} and a SVC pump would further increase the pressure in the opposing venous territory (IVC); and 2) the majority of the systemic venous return is carried by the IVC.

Trusty et al.~\cite{trusty2019vitro,Trusty2019} studied installation of VentrioFlo Tru Pulse Pump, PediMag and CentriMag in parallel with the Fontan baffle and observed significant recirculation of the blood flow through the Fontan pathway. The majority of the flow recirculated from the outflow cannula, down through the Fontan baffle and back into the inflow cannula of the pump. Trusty et al.~\cite{Trusty2019} argued that the hemodynamic influence of the recirculation phenomenon depends on the amount of the device’s flow output. They showed that to achieve hemodynamic improvement using the PediMag device, it is essential to constrict the Fontan baffle to limit the blood flow recirculation through the pump; however, this resulted in a detrimental increase in the SVC pressure. Whereas, while using the CentriMag pump (a stronger device with greater output) the recirculation was beneficial for 3 reasons: 1) there was no need for banding the Fontan baffle 2) there was no increase in the SVC pressure, and 3) the recirculation actually entrained both the IVC and SVC flow which allowed both vena cavae pressures to decrease while increasing pulmonary arterial pressure.

Nevertheless, banding of the Fontan baffle or recirculation caused by the capvopulmonary pump could significantly increase thrombosis risks~\cite{honjo2019commentary}, therefore not a desirable option. For a series device implementation, cavopulmonary support-associated thrombosis could result in sudden death of a patient and therefore post-operative anti-thrombotic medications are crucial in clinical management of these patients."

\begin{figure}[tb!]
    \centerline{ \includegraphics[width=88MM]{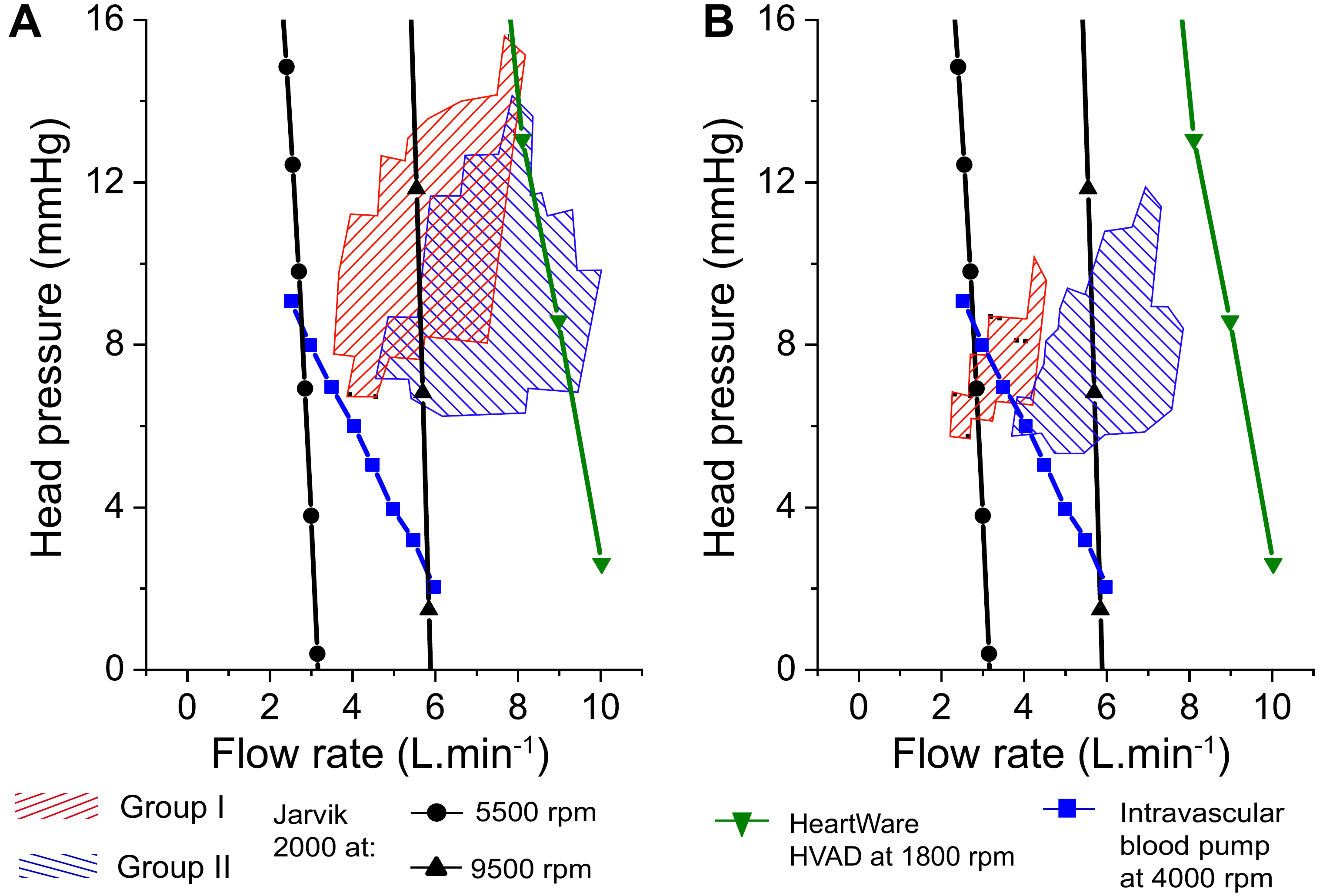}}
    \caption{\textbf{Evaluating off-label use of available VADs (or proposed cavopulmonary device prototypes) for right-side support application.} Desired operating regions of (\textbf{A}) full and (\textbf{B}) IVC assist cavopulmonary devices overlaid with characteristics curves of a left VAD and a cavopulmonary intravascular blood pump.}
         \label{fig6}
\end{figure}

The results of our current study can also be used to evaluate the performance of the available cavopulmonary prototype devices as well as off-label use of commercially available left VADs for cavopulmonary applications. Researchers in recent years have investigated several cavopulmonary devices for supporting the failing Fontan circulation. Throckmorton et al.~\cite{throckmorton2011numerical} proposed an intravascular blood pump for the IVC support configuration. As the intravascular blood pump (at 4000rpm) generates a pressure increase of 2.5-9.5mmHg for the flows rates of 2-6L.min$^{-1}$, its performance curve covers parts of the IVC support desired operating regions (Fig.~\ref{fig6}) which suggests that this intravascular cavopulmonary blood pump at 4000rpm can serve some of the failing patients.
Fig.~\ref{fig6} and Table~\ref{table2} illustrate how the desired operating regions reported in this study can help evaluate experimental devices and the off-label use of commercially available left VADs such as the Jarvik 2000 VAD and HeartWare HVAD for cavopulmonary support.

\begin{table}[tb!]
\centering
\caption{Example assessment of off-label use of available left VADs and a proposed cavopulmonary device prototypes for right-side support application based on Fig.~\ref{fig6}.}
\begin{tabular}{lcc}
\hline
\textbf{}                                                                                                                                     & \multicolumn{2}{c}{\textbf{Likely beneficial for}} \\ \cline{2-3} 
                                                                                                                                              & \textbf{Full support?}   & \textbf{IVC support?}   \\ \hline
\textbf{\begin{tabular}[c]{@{}l@{}}Jarvik 2000 \\ at 5500rpm~\cite{farahmand2019risks}\end{tabular}}                   & No                     & Yes, only group I                         \\ \hline
\textbf{\begin{tabular}[c]{@{}l@{}}Jarvik 2000 \\ at 9500rpm\end{tabular}}                                                                    & Yes, group I and II    & Yes, only group II                        \\ \hline
\textbf{\begin{tabular}[c]{@{}l@{}}HeartWare HVAD\\ at 1800rpm~\cite{moazami2013axial}\end{tabular}}                   & Yes, only group II     & No                                        \\ \hline
\textbf{\begin{tabular}[c]{@{}l@{}}Intravascular blood pump\\ at 4000rpm~\cite{throckmorton2011numerical}\end{tabular}} & No                     & Yes, only group I                         \\ \hline
\end{tabular}
\label{table2}
\end{table}

\textit{Limitations}

Factors such as the TCPC configuration and geometry have influence on the target device operation. We fully recognize that differences in the surgical junction geometry can affect the dynamic resistance and the obtained cavopulmonary support operating regions. Further studies are needed to quantify the flow resistance in different implementation of the full support surgical junction geometry. Furthermore, the geometry of the TCPC junction can affect the hepatic flow distribution~\cite{yang2012hepatic,jagani2019dual} which is not captured by the LPN model. Computational fluid dynamic simulations can be performed to carefully investigate the change in the hepatic flow distribution as the result of different Fontan geometries in the presence of a cavopulmonary assist.

Since cavopulmonary support is not currently employed in the clinical management of Fontan patients, clinical data does not exist to validate the results of this study. We believe that cavopulmonary support will result in positive remodeling of the vasculature leading to a reduction in pulmonary vascular resistance. However, the goal of our current models is to focus on acute physiologic outcomes rather than long-term adaptation of the pulmonary and systemic vascular resistances as a result of the cavopulmonary support. Our models also do not include fenestration that is present in some of the patients. However, the fenestration most likely would be surgically removed if cavopulmonary support is to be implemented.
\section{Conclusion}
This study presents the hydraulic operating regions for manufacturing a cavopulmonary blood pump specifically designed for helping failing Fontan patients. We simulated and investigated the interaction of a cavopulmonary assist device with a full range of failing stage Fontan physiologies reflecting pathological conditions with various severity levels of systolic and diastolic dysfunction and abnormal pulmonary and systemic vascular resistances, representing approximately 95\% of the adult failing Fontan population. 

The results show that a cavopulmonary assist device could increase cardiac index by 35\% and decrease the IVC pressure up to 45\% depending on the patient’s pre-support hemodynamic state and surgical configuration of the cavopulmonary assist device (IVC or full support). Overall, according to the identified desired operating region, the IVC assist configurations is not suitable for a large fraction of the traditional group of failing Fontan patients (group I). For the newly recognized failing class of Fontan patients with normal cardiac output and high IVC pressure (group II) both IVC and full assist support can be beneficial. 

For group II, an IVC assist device should be able to create head pressures of 5-12mmHg for flows of $\approx$3.5-7.5L.min$^{-1}$. However, the cavopulmonary device should cover a wider range for both groups when providing full support.

Additionally, we identified physiologic scenarios where cavopulmonary support cannot benefit hemodynamic conditions, but rather predisposes these physiologic scenarios to perfusion lung damage or venous collapse as a result of the pressure generated by the cavopulmonary device. These physiologic scenarios were excluded and were not used for obtaining the desired operating regions. These results may help serve as guidelines for identifying patient characteristics unsuitable for cavopulmonary support.

\appendices
\section{Total Resistance of the IVC Support Surgical Junction}

\subsection{Introduction}
Implementation of the cavopulmonary support in Fontan patients will result in increased blood flow through the cavopulmonary connection. Previous researches have shown that the caval flow collision in a TCPC junction results in significant energy loss [1]. The increases in this energy loss as the result of augmented flow induced by cavopulmonary support may be significant and should be included in the simulations to obtain more accurate results. Therefore, we characterized this resistance at different flow rates using a physical experiment containing the IVC support surgical junction geometry.

\subsection{IVC support surgical junction geometry}
We constructed an idealized IVC surgical junction geometry (Fig.~\ref{Fig.S1}) and physically reproduced the corresponding flow phantom using a high resolution 3D printer (Connex 350 PolyJet, Stratasys, Inc., Eden Prairie, MN) and rigid material (VeroClear, Stratasys, Inc., Eden Prairie, MN). We performed steady flow tests, data acquisition, and signal processing as described below. 
\setcounter{figure}{0}
\renewcommand\thefigure{A.\arabic{figure}}
\begin{figure}[tb!]
    \centerline{\includegraphics[width=60mm]{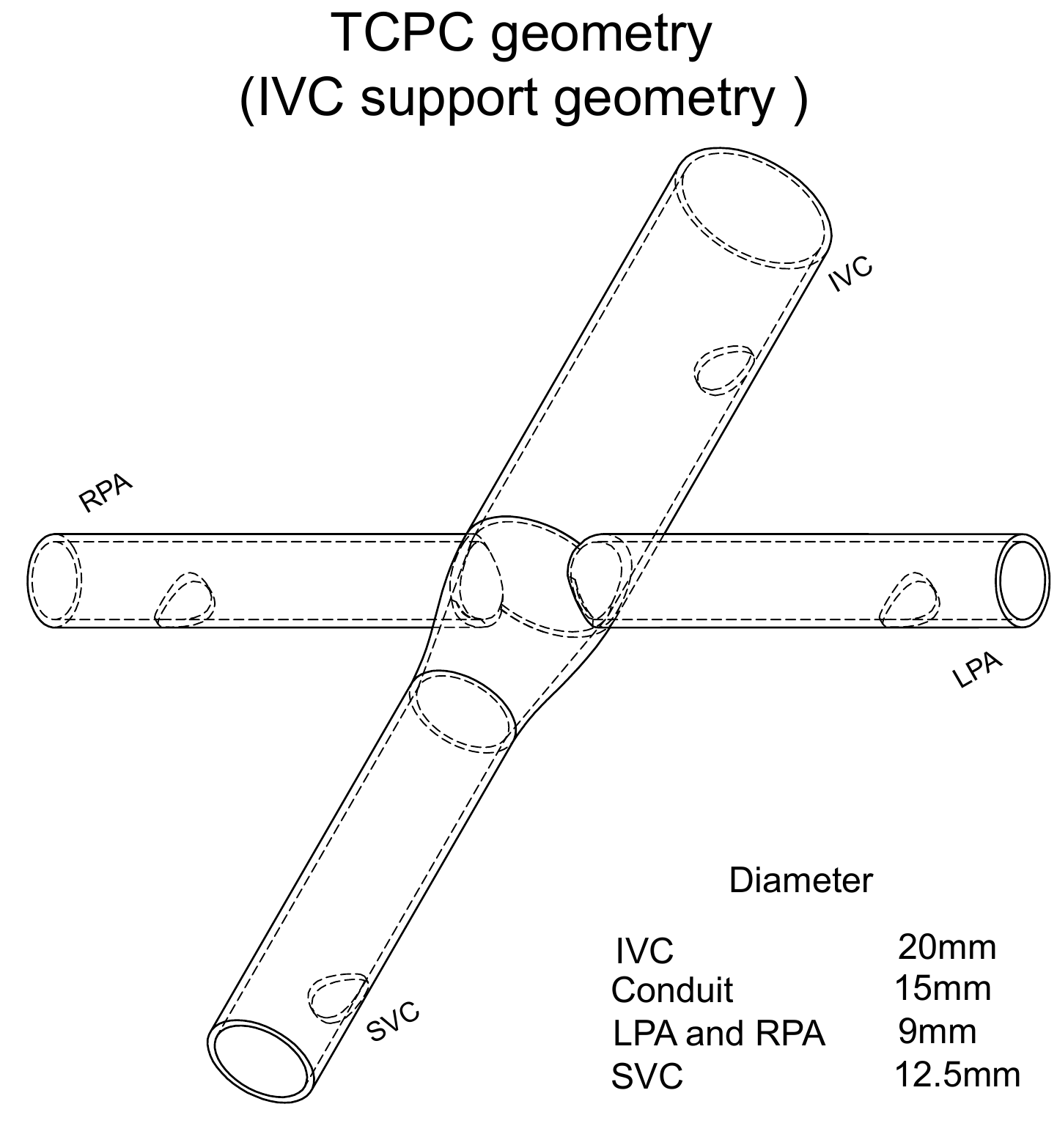}}
    \caption{Geometry of physical phantoms. \textbf{LPA} is left pulmonary artery, \textbf{RPA} is right pulmonary artery \textbf{SVC} is superior vena cava, \textbf{IVC} is inferior vena cava.}
     \label{Fig.S1}
\end{figure}

\begin{figure}[tb!]
    \centerline{\includegraphics[width=88mm]{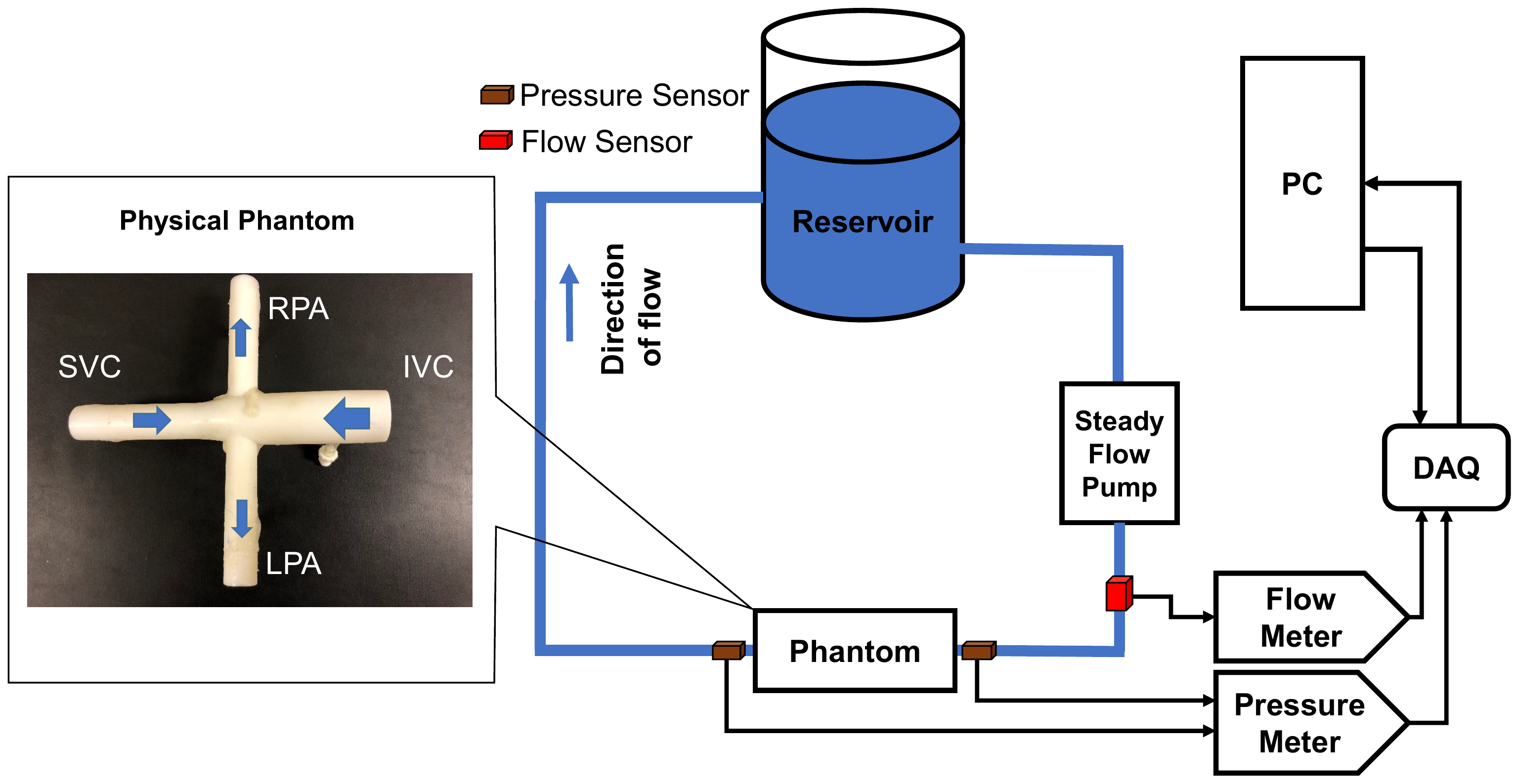}}
    \caption{physical hydraulic experiment setup. Photograph of IVC support surgical junction geometry and schematic of the physical experiment setup.}
     \label{Fig.S2}
\end{figure}

\subsection{Experimental setup (Fig.~\ref{Fig.S2})}
We performed steady flow tests on the flow phantoms, data acqusition, and signal processsing as described here. We used a glycerol solution (60\% water and 40\% glycerol) as the working fluid in the setup~\cite{kung2019hybrid,Mirzaei2020}. The density of the fluid and dynamic viscosity of the working fluid are 1092.4 kg.m$^{-3}$ and 0.0041 Pa.s, respectively; similar to those of blood. A steady-flow pump was used to drive the flow through the flow loop and the flow rate through each inlet and outlet of the phantom was controlled using a mechanical valve. Volumetric flow rate was measured using a external clamp on ultra sound flow sensor (12PXL, Transonic systems, NY). The pressure at different locations were measured using catheter pressure transducers attached to a pressure unit conditioner (PCU2000, Millar Instruments, TX). We used separate NI modules for data acquisition (NI 9205, NI 9263, National Instruments, TX). We collected the data at a sampling frequency of 1000 HZ. All signal were passed though a 40 Hz low-pass filter for noise canceling. We measured the IVC resistance at different cardiac outputs by comparing the flow rates through each geometry with the corresponding pressure changes ($\Delta$P = R$\times$Q, where R is resistance, P is pressure, and Q is flow). Then, we fitted a linear line with adequate goodness-of-fit (\textit{r$^{2}$}=0.99) to correlate the resistance values with the total flow rate (i.e. cardiac output) through the phantom.

\subsection{Results}
The estimated total resistance (R [$\frac{mmHg}{\frac{L}{min}}$]) and pressure loss ($\Delta$P$_{loss}$ [mmHg]) across the IVC support surgical geometry as a function of the cardiac output (CO [$\frac{L}{min}$]) are as follows:\\\\
R  = 0.1234 $\times$ CO + 0.1101 \\
$\Delta$P$_{loss}$ = CO$\times$R \\

This pressure loss estimate at different flow rates is in excellent agreement with the data reported by Sundareswaran et al.~\cite{sundareswaran2008total}.
\section{Flow Diagram illustrating the procedure for obtaining a desired operating region (Fig.~\ref{figA1})}
\setcounter{figure}{0}
\renewcommand\thefigure{B.\arabic{figure}}
\begin{figure}[tbh!]
 \centerline{\includegraphics[width=88mm]{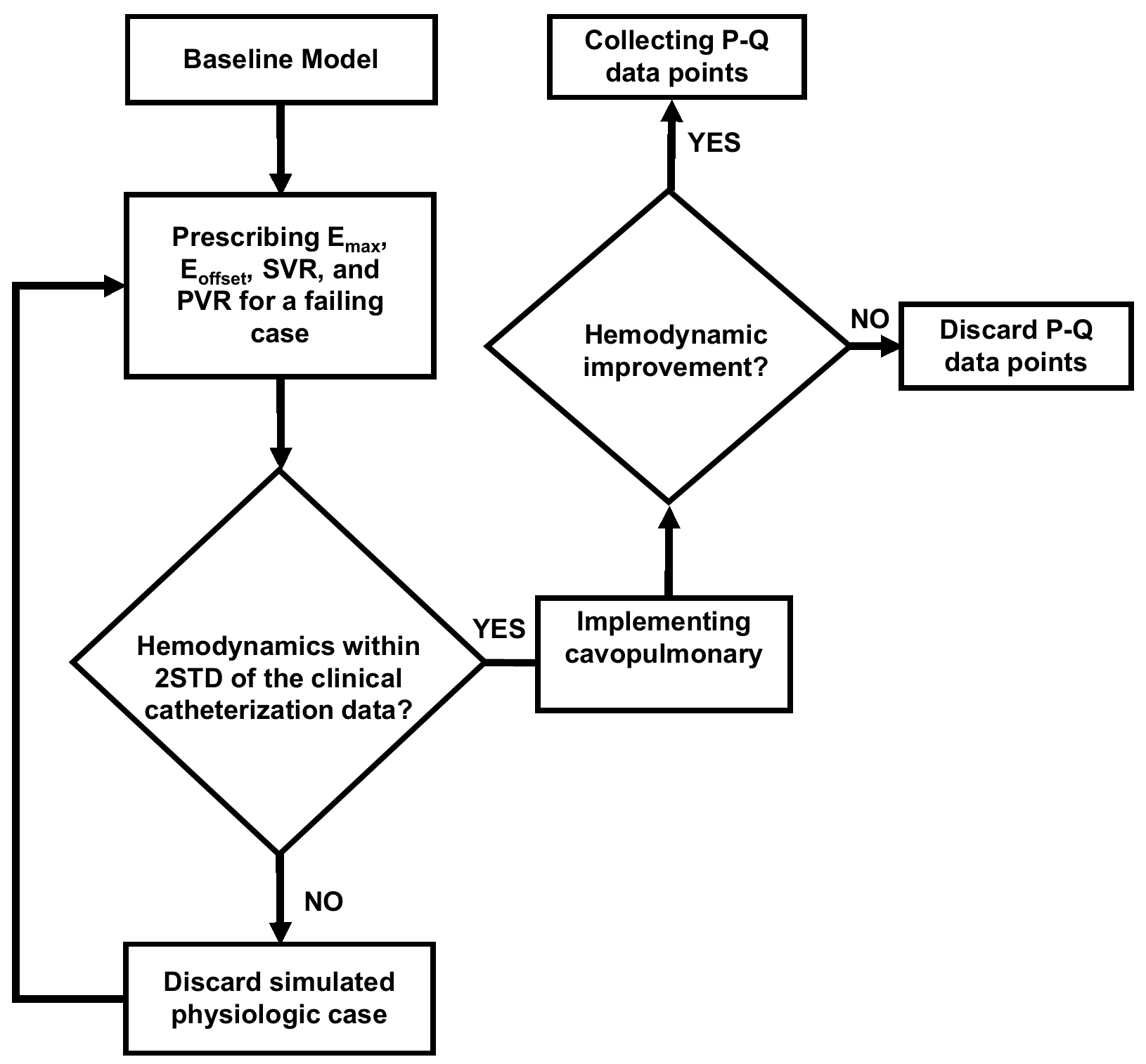}}\label{appfig1}
    \caption{Flow Diagram illustrating the procedure for obtaining a desired operating region.}
         \label{figA1}
\end{figure}
\section*{Acknowledgment}

The authors acknowledge the work of members of our laboratories, especially Daniel Custer for his assistance in manuscript editing. 
\ifCLASSOPTIONcaptionsoff
  \newpage
\fi

\bibliographystyle{IEEEtran}
\bibliography{references}

\begin{thebibliography}{10}
\providecommand{\url}[1]{#1}
\csname url@samestyle\endcsname
\providecommand{\newblock}{\relax}
\providecommand{\bibinfo}[2]{#2}
\providecommand{\BIBentrySTDinterwordspacing}{\spaceskip=0pt\relax}
\providecommand{\BIBentryALTinterwordstretchfactor}{4}
\providecommand{\BIBentryALTinterwordspacing}{\spaceskip=\fontdimen2\font plus
\BIBentryALTinterwordstretchfactor\fontdimen3\font minus
  \fontdimen4\font\relax}
\providecommand{\BIBforeignlanguage}[2]{{%
\expandafter\ifx\csname l@#1\endcsname\relax
\typeout{** WARNING: IEEEtran.bst: No hyphenation pattern has been}%
\typeout{** loaded for the language `#1'. Using the pattern for}%
\typeout{** the default language instead.}%
\else
\language=\csname l@#1\endcsname
\fi
#2}}
\providecommand{\BIBdecl}{\relax}
\BIBdecl

\bibitem{davey2019surveillance}
B.~T. Davey \emph{et~al.}, ``Surveillance and screening practices of new
  england congenital cardiologists for patients after the fontan operation,''
  \emph{Congenital heart disease}, 2019.

\bibitem{van2018state}
J.~P. van~der Ven \emph{et~al.}, ``State of the art of the fontan strategy for
  treatment of univentricular heart disease,'' \emph{F1000Research}, vol.~7,
  2018.

\bibitem{rodefeld2011cavopulmonary}
M.~D. Rodefeld \emph{et~al.}, ``Cavopulmonary assist:(em) powering the
  univentricular fontan circulation,'' in \emph{Seminars in Thoracic and
  Cardiovascular Surgery: Pediatric Cardiac Surgery Annual}, vol.~14,
  no.~1.\hskip 1em plus 0.5em minus 0.4em\relax Elsevier, 2011, pp. 45--54.

\bibitem{kuwabara2018liver}
M.~Kuwabara \emph{et~al.}, ``Liver cirrhosis and/or hepatocellular carcinoma
  occurring late after the fontan procedure―a nationwide survey in
  japan―,'' \emph{Circulation Journal}, vol.~82, no.~4, pp. 1155--1160, 2018.

\bibitem{ohuchi2017hemodynamic}
H.~Ohuchi \emph{et~al.}, ``Hemodynamic determinants of mortality after fontan
  operation,'' \emph{American heart journal}, vol. 189, pp. 9--18, 2017.

\bibitem{GEWILLIG2014105}
M.~Gewillig and D.~J. Goldberg, ``Failure of the fontan circulation,''
  \emph{Heart Failure Clinics}, vol.~10, no.~1, pp. 105 -- 116, 2014, heart
  Failure in Adult Congenital Heart Disease.

\bibitem{ohuchih2014comparisonof}
H.~Ohuchi \emph{et~al.}, ``Comparison of prognostic variables in children and
  adults with fontan circulation,'' \emph{International journal of cardiology},
  vol. 173, no.~2, pp. 277--283, 2014.

\bibitem{book2016clinical}
W.~M. Book \emph{et~al.}, ``Clinical phenotypes of fontan failure: implications
  for management,'' \emph{Congenital heart disease}, vol.~11, no.~4, pp.
  296--308, 2016.

\bibitem{ohuchi2017optimal}
H.~Ohuchi, ``Where is the “optimal” fontan hemodynamics?'' \emph{Korean
  circulation journal}, vol.~47, no.~6, pp. 842--857, 2017.

\bibitem{miranda2019haemodynamic}
W.~R. Miranda \emph{et~al.}, ``Haemodynamic profiles in adult fontan patients:
  associated haemodynamics and prognosis,'' \emph{European journal of heart
  failure}, 2019.

\bibitem{goldberg2017hepatic}
D.~J. Goldberg \emph{et~al.}, ``Hepatic fibrosis is universal following fontan
  operation, and severity is associated with time from surgery: a liver biopsy
  and hemodynamic study,'' \emph{Journal of the American Heart Association},
  vol.~6, no.~5, p. e004809, 2017.

\bibitem{trusty2018impact}
P.~M. Trusty \emph{et~al.}, ``Impact of hemodynamics and fluid energetics on
  liver fibrosis after fontan operation,'' \emph{The Journal of thoracic and
  cardiovascular surgery}, vol. 156, no.~1, pp. 267--275, 2018.

\bibitem{de1998fontan}
M.~R. de~Leval, ``The fontan circulation: What have we learned? what to
  expect?'' \emph{Pediatric cardiology}, vol.~19, no.~4, pp. 316--320, 1998.

\bibitem{Trusty2019}
P.~Trusty \emph{et~al.}, ``{An in vitro analysis of the PediMag and CentriMag
  for right-sided failing Fontan support},'' \emph{The Journal of Thoracic and
  Cardiovascular Surgery}, vol. 158, no.~5, pp. 1413--1421, nov 2019.

\bibitem{kung2014simulation}
E.~Kung \emph{et~al.}, ``A simulation protocol for exercise physiology in
  fontan patients using a closed loop lumped-parameter model,'' \emph{Journal
  of biomechanical engineering}, vol. 136, no.~8, p. 081007, 2014.

\bibitem{cavalcanti2001analysis}
S.~Cavalcanti \emph{et~al.}, ``Analysis by mathematical model of haemodynamic
  data in the failing fontan circulation,'' \emph{Physiological measurement},
  vol.~22, no.~1, p. 209, 2001.

\bibitem{bryc2012normal}
W.~Bryc, \emph{The Normal Distribution: Characterizations with Applications},
  ser. Lecture Notes in Statistics.\hskip 1em plus 0.5em minus 0.4em\relax
  Springer New York, 2012.

\bibitem{di2015simulation}
A.~Di~Molfetta \emph{et~al.}, ``Simulation of ventricular, cavo-pulmonary, and
  biventricular ventricular assist devices in failing f ontan,''
  \emph{Artificial organs}, vol.~39, no.~7, pp. 550--558, 2015.

\bibitem{montani2013pulmonary}
D.~Montani \emph{et~al.}, ``Pulmonary arterial hypertension,'' \emph{Orphanet
  journal of rare diseases}, vol.~8, no.~1, p.~97, 2013.

\bibitem{larue2017clinical}
S.~J. LaRue \emph{et~al.}, ``Clinical outcomes associated with
  intermacs-defined right heart failure after left ventricular assist device
  implantation,'' \emph{The Journal of Heart and Lung Transplantation},
  vol.~36, no.~4, pp. 475--477, 2017.

\bibitem{gewillig2016fontan}
M.~Gewillig and S.~C. Brown, ``The fontan circulation after 45 years: update in
  physiology,'' \emph{Heart}, vol. 102, no.~14, pp. 1081--1086, 2016.

\bibitem{pretre2008right}
R.~Pr{\^e}tre \emph{et~al.}, ``Right-sided univentricular cardiac assistance in
  a failing fontan circulation,'' \emph{The Annals of thoracic surgery},
  vol.~86, no.~3, pp. 1018--1020, 2008.

\bibitem{broda2018progress}
C.~Broda \emph{et~al.}, ``Progress in experimental and clinical subpulmonary
  assistance for fontan circulation,'' \emph{The Journal of thoracic and
  cardiovascular surgery}, vol. 156, no.~5, pp. 1949--1956, 2018.

\bibitem{zhou2019avalonelite}
C.~Zhou \emph{et~al.}, ``Avalonelite double lumen cannula for total
  cavopulmonary assist in failing fontan sheep model with valved extracardiac
  conduit,'' \emph{Asaio Journal}, vol.~65, no.~4, pp. 361--366, 2019.

\bibitem{lin2019computational}
W.~P. Lin \emph{et~al.}, ``Computational fluid dynamic simulations of a
  cavopulmonary assist device for failing fontan circulation,'' \emph{The
  Journal of thoracic and cardiovascular surgery}, 2019.

\bibitem{pekkan2018vitro}
K.~Pekkan \emph{et~al.}, ``In vitro validation of a self-driving aortic-turbine
  venous-assist device for fontan patients,'' \emph{The Journal of thoracic and
  cardiovascular surgery}, vol. 156, no.~1, pp. 292--301, 2018.

\bibitem{shimizu2016partial}
S.~Shimizu \emph{et~al.}, ``Partial cavopulmonary assist from the inferior vena
  cava to the pulmonary artery improves hemodynamics in failing fontan
  circulation: a theoretical analysis,'' \emph{The Journal of Physiological
  Sciences}, vol.~66, no.~3, pp. 249--255, 2016.

\bibitem{riemer2005mechanical}
R.~K. Riemer \emph{et~al.}, ``Mechanical support of total cavopulmonary
  connection with an axial flow pump,'' \emph{The Journal of thoracic and
  cardiovascular surgery}, vol. 130, no.~2, pp. 351--354, 2005.

\bibitem{rodefeld2004cavopulmonary}
M.~D. Rodefeld \emph{et~al.}, ``Cavopulmonary assist in the neonate: an
  alternative strategy for single-ventricle palliation,'' \emph{The Journal of
  thoracic and cardiovascular surgery}, vol. 127, no.~3, pp. 705--711, 2004.

\bibitem{farahmand2019risks}
M.~Farahmand \emph{et~al.}, ``Risks and benefits of using a commercially
  available ventricular assist device for failing fontan cavopulmonary support:
  A modeling investigation,'' \emph{IEEE Transactions on Biomedical
  Engineering}, 2019.

\bibitem{jagani2019dual}
J.~N. Jagani \emph{et~al.}, ``Dual-propeller cavopulmonary pump for assisting
  patients with hypoplastic right ventricle.'' \emph{ASAIO journal (American
  Society for Artificial Internal Organs: 1992)}, 2019.

\bibitem{trusty2019vitro}
P.~M. Trusty \emph{et~al.}, ``In vitro examination of the ventriflo true pulse
  pump for failing fontan support,'' \emph{Artificial organs}, vol.~43, no.~2,
  pp. 181--188, 2019.

\bibitem{honjo2019commentary}
O.~Honjo \emph{et~al.}, ``Commentary: Engineering an optimal mechanical
  circulatory support system for the cavopulmonary connection.'' \emph{The
  Journal of thoracic and cardiovascular surgery}, 2019.

\bibitem{throckmorton2011numerical}
A.~L. Throckmorton \emph{et~al.}, ``Numerical, hydraulic, and hemolytic
  evaluation of an intravascular axial flow blood pump to mechanically support
  fontan patients,'' \emph{Annals of biomedical engineering}, vol.~39, no.~1,
  pp. 324--336, 2011.

\bibitem{moazami2013axial}
N.~Moazami \emph{et~al.}, ``Axial and centrifugal continuous-flow rotary pumps:
  a translation from pump mechanics to clinical practice,'' \emph{The Journal
  of Heart and Lung Transplantation}, vol.~32, no.~1, pp. 1--11, 2013.

\bibitem{yang2012hepatic}
W.~Yang \emph{et~al.}, ``Hepatic blood flow distribution and performance in
  conventional and novel y-graft fontan geometries: a case series computational
  fluid dynamics study,'' \emph{The Journal of thoracic and cardiovascular
  surgery}, vol. 143, no.~5, pp. 1086--1097, 2012.

\bibitem{kung2019hybrid}
E.~Kung \emph{et~al.}, ``A hybrid experimental-computational modeling framework
  for cardiovascular device testing,'' \emph{Journal of biomechanical
  engineering}, vol. 141, no.~5, p. 051012, 2019.

\bibitem{Mirzaei2020}
\BIBentryALTinterwordspacing
E.~Mirzaei \emph{et~al.}, ``An algorithm for coupling multibranch in vitro
  experiment to numerical physiology simulation for a hybrid cardiovascular
  model,'' \emph{International Journal for Numerical Methods in Biomedical
  Engineering}, p. e3289. [Online]. Available:
  \url{https://onlinelibrary.wiley.com/doi/abs/10.1002/cnm.3289}
\BIBentrySTDinterwordspacing

\bibitem{sundareswaran2008total}
K.~S. Sundareswaran \emph{et~al.}, ``The total cavopulmonary connection
  resistance: a significant impact on single ventricle hemodynamics at rest and
  exercise,'' \emph{American Journal of Physiology-Heart and Circulatory
  Physiology}, vol. 295, no.~6, pp. H2427--H2435, 2008.

\end{thebibliography}
\end{document}